\documentclass[prl,twocolumn,amsmath,amssymb,floatfix,reprint]{revtex4-1}
\usepackage{graphicx}
\usepackage{dcolumn}
\usepackage{bm}
\usepackage{color}
\usepackage[usenames,dvipsnames]{xcolor}
\usepackage{mciteplus}
\usepackage{braket}
\usepackage{ulem}

\usepackage[utf8]{inputenc}
\usepackage[T1]{fontenc}
\usepackage{amsmath}
\usepackage{siunitx}
\usepackage{cleveref}

\begin{document}
\title{Turbulent properties of stationary flows in porous media}

\author{Florencia Falkinhoff$^{1,2,3}$,
  Alexandre Ponomarenko$^{2}$, Jean-Lou Pierson$^{1}$, Lionel Gamet$^{1}$, Romain Volk$^{2}$ and Micka\"{e}l Bourgoin$^{2}$}

\affiliation{$^{1}$IFP Energies Nouvelles, 69360, Solaize, France \\
$^{2}$Ens de Lyon, CNRS, Laboratoire de physique, F-69342 Lyon, France \\
$^{3}$Max Planck Institute for Dynamics and Self-Organization, G\"{o}ttingen, Germany
}

\date{\today}

\begin{abstract}
In this study, we investigated the flow dynamics in a fixed bed of hydrogel beads using Particle Tracking Velocimetry to compute the velocity field in the middle of the bed for moderate Reynolds numbers. We discovered that despite the overall stationarity of the flow and relatively low Reynolds number, it exhibits complex multiscale spatial dynamics reminiscent of those observed in classical turbulence. We found evidence of the presence of an inertial range and a direct energy cascade, and were able to obtain a value for a "porous" Kolmogorov constant of $C_2 = 3.1\pm 0.3$. This analogy with turbulence opens up new possibilities for understanding mixing and global transport properties in porous media.
\end{abstract}

\maketitle

Flows in porous media are crucial in an immense variety of natural and industrial systems. These include for instance oil and gas extraction from underground wells through porous rocks, cooling of nuclear plants, and chemical reactions such as catalysis in packed bed reactors and separation processes~\citep{Wehinger2015,trogadas2016}. They also play a major role in new clean energy developments such as harnessing geothermal energy from underground reservoirs~\citep{banks2018} and hydrogen storage~\citep{chen2022}, to biological flows and bioengineering applications~\citep{Jensen2016, Peyrounette2018,helder2012}, etc.. 

\begin{figure}[t]
  \centerline{
    \includegraphics[width=0.75\columnwidth,trim={3.9cm 0.2cm 4.5cm 0},clip]{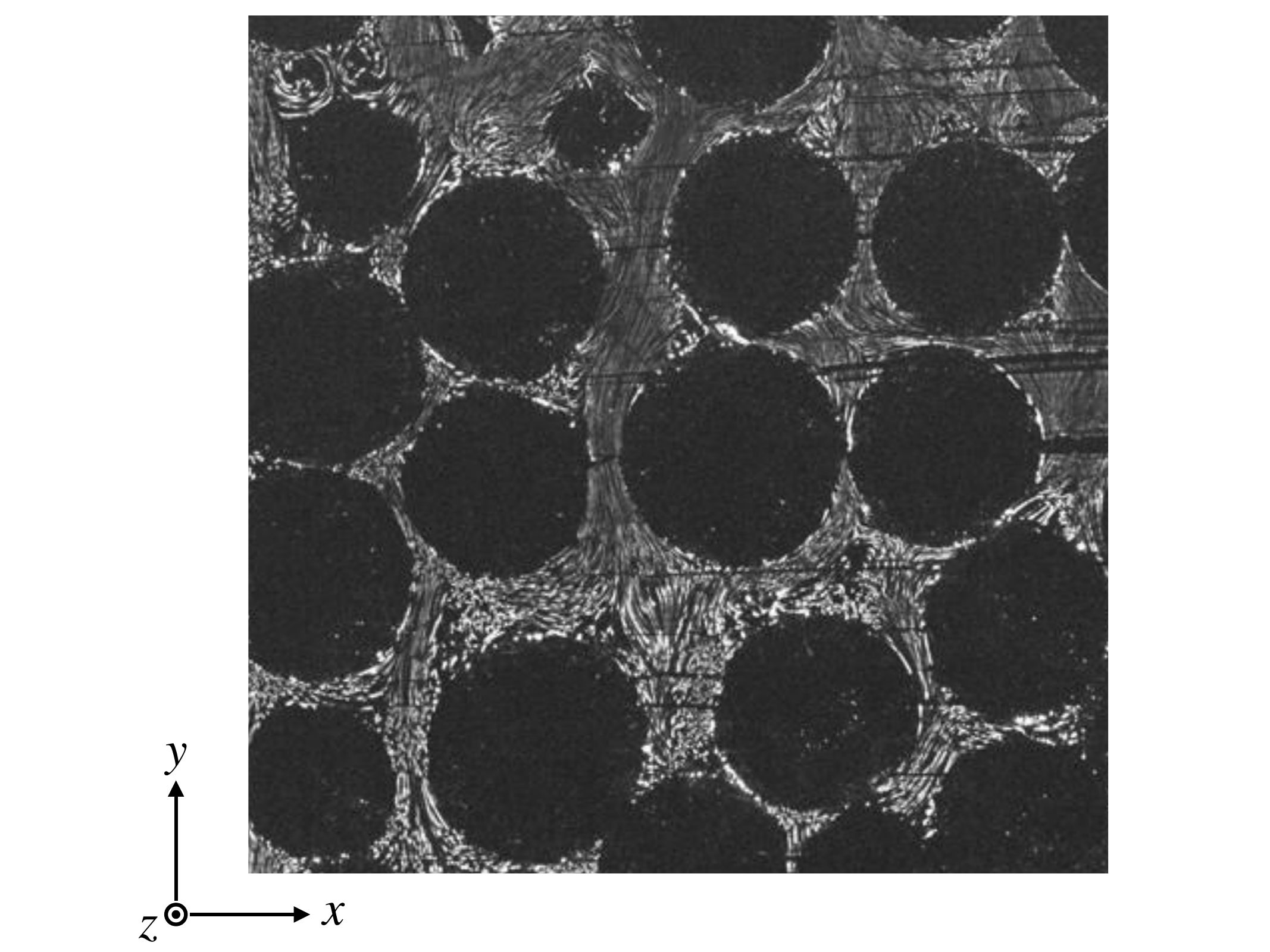}}
  \caption{Long-time exposure of a generic experiment. The tracer particles are shown in greys and the hydrogel beads are the empty spaces in the middle. The Lagrangian dynamics is not trivial and is observed at the local scale. It is reflective of the complex nature of the system.}
 \label{fig:snapshot}
\end{figure}

Understanding the hydrodynamics of flows in such situations, and its impact on mass and heat transport phenomena, is therefore of primary importance in many scientific fields and applications. The question is particularly complex when the flow in the porous bed is intense (namely when its Reynolds number, defined below, is significantly larger than one) so that inertial contributions to its dynamics cannot be neglected. This results in the emergence of a variety of intricate structures~\citep{Patil2013,apte2022_JFM} developing at the scale of each pore (see fig.~\ref{fig:snapshot}) and impacting the global transport properties at play. At first sight, this complexity shares qualitative similarities with the properties of turbulence in homogeneous fluids, where the multiscale dynamics is crucial to the mixing and transport efficiency of turbulent flows. 
Qualitative analogies with turbulence were even stressed regarding the Lagrangian dynamics of tracer particles in porous medium flows at low Reynolds number. Namely, Holzner~\textit{et al.}~\cite{Holzner2015} reported highly non-Gaussian Lagrangian acceleration statistics in a porous medium flow at Reynolds number $Re < 1$, strikingly similar to those observed in turbulent flows~\cite{Voth2001,LaPorta2001}.

Figure~\ref{fig:snapshot} shows a typical long time exposure image of the flow recorded in our experiment (details will be given below). The flow shows a clear spatial complexity, with shear zones and strain, rotation dominated structures, stagnation points, etc., which recall the characteristic multiscale structures of a turbulent flow. Even if the flow remains steady in time, as shown in the supplementary material~\cite{suppMat}, this spatial multiscale behavior is reminiscent of the energy cascade phenomenology in turbulent flows. A direct energy cascade in turbulence occurs when the mechanical energy injected into a flow is transferred from the injection scale ($L$) down to the viscous dissipative scale ($\eta$), thus resulting in the multiscale nature of the system. The range of scales $\eta\ll r\ll L$ is the so-called \textit{inertial range} of turbulence. Following Kolmogorov's phenomenology (hereafter referred to as K41)~\citep{bib:k41,frisch1995} where this process is formulated in a self-similar description, the random multiscale dynamics of turbulence is classically described in terms of the Eulerian velocity structure functions, defined as the statistical moments of the velocity increments, $\delta_r \mathbf{u}=\mathbf{u}(\mathbf{x}+\mathbf{r},t)-\mathbf{u}(\mathbf{x},t)$
between points of the flow separated by a distance $r=|\mathbf{r}|$. K41 predicts that within inertial scales, the structure functions should scale as $S_p (r)= \langle |\delta_r \mathbf{u}|^p\rangle \propto (\epsilon r)^{p/3}$, for $r$ within the inertial range and $\epsilon$ is the energy injection/dissipation rate per unit mass. Here, we will focus on second ($S_2$) and third ($S_3$) order statistics, whose physical relevance is fundamental as they relate respectively to the distribution of energy across scales and to the direction of the energy flux across scales.

The main goal of this Letter is to explore to which extent the spatial complexity of this steady porous medium flow shares quantitative similarities with turbulence. For that purpose we analyse  transitional flows (i.e., flows that are neither laminar nor turbulent, see~\cite{Wood2020}) in porous media with the statistical tools of turbulence to characterize the spatial fluctuations of velocity and their correlations. Our experiments and analysis reveal striking similarities with ``classical'' fluid turbulence, as the multiscale Eulerian dynamics extracted from the porous media flow are found to be indistinguishable form that of homogeneous fluid  turbulence.

In order to explore the hydrodynamics in the core of a porous medium, we studied the local flow in a fixed bed made of spherical particles contained in a cylinder of diameter $D=9$~cm and height $H=40$~cm. To be able to measure inside the pores we deployed state-of-the art Particle Tracking Velocimetry (PTV) using index-matched hydrogel beads of mean diameter $d=1.4$~cm to make the porous bed, where small (30~$\mu$m in diameter) fluorescent tracer particles are seeded into the fluid and tracked in 3D with two high-speed cameras.

\begin{figure}
  \centerline{
    \includegraphics[width=8.8cm,clip]{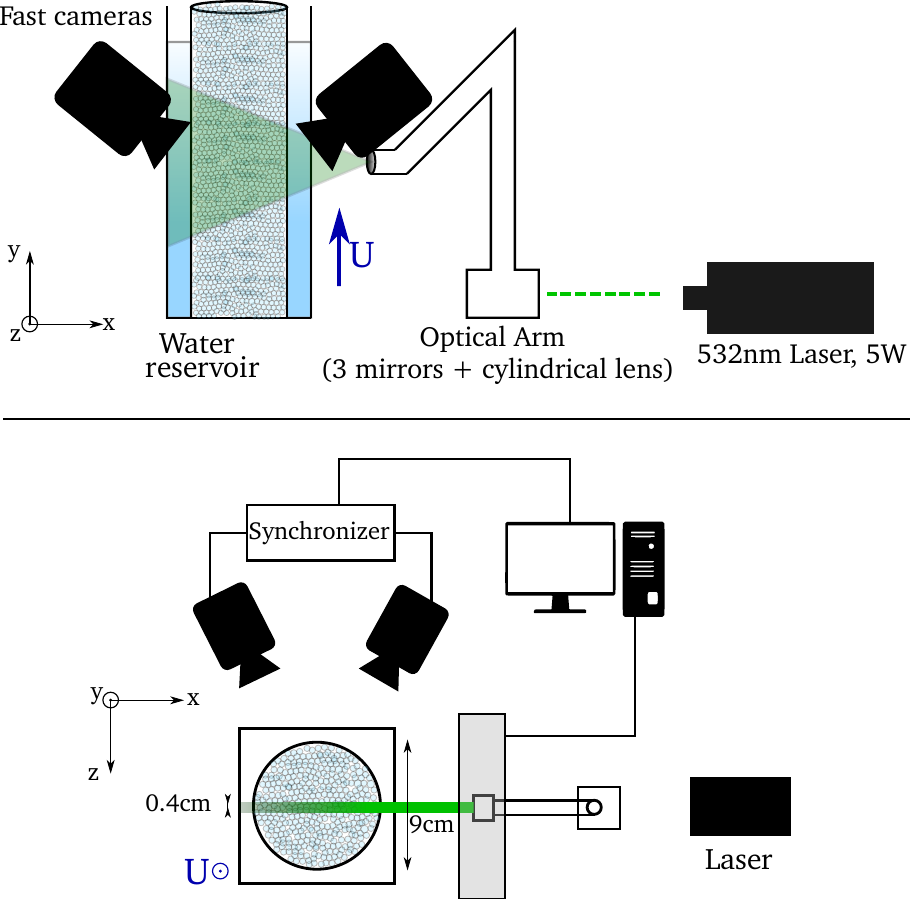}}
  \caption{Schematic representation of the experimental setup.}
 \label{fig:setup}
\end{figure}

 More specifically, the setup, schematized in figure \ref{fig:setup}, consists of a closed water loop circuit and the porous bed of hydrogel beads is fixed with the aid of two grids at the top and bottom of the test section preventing the fluidization of the bed. A centrifugal pump is used to drive the flow of water, at a constant flow rate $Q$ which can be accurately prescribed with a solenoid valve. The flow rate is monitored by a magnetic flow meter, which provides a direct measurement of the mean superficial velocity through the bed, $U=4Q/(\pi D^2)$. This is used to define the Reynolds number $Re=Ud/\nu$ of the flow, where $\nu=10^{-6}$m$^2$.s$^{-1}$ is the kinematic viscosity of water. We studied the local flow at four different Reynolds numbers: $Re=[124,169,203,211]$. The flow at these Reynolds is found to remain steady without any signature of temporal fluctuations \cite{Wood2020}. The saturating water is seeded with $31\mu$m tracer particles whose motion is recorded by two \textit{Phantom\ v12} high-speed cameras, which were used to record several 2-second films at 2200 fps  and using a 12-bit, 880 $\times$ 896 px resolution. The bed is illuminated by a 5W laser with a 532nm wavelength, shaped into a thick fixed sheet parallel to the flow rate generated by a cylindrical lens. This gives a visualization region of approximately $(5\times5\times0.4)$cm (length$\times$width$\times$depth), which in terms of the hydrogel diameter is $(3.6\times 3.6\times 0.3)d$. As shown in figure \ref{fig:setup}, the flow is driven in the $y-$ direction, and the $xz$-plane is perpendicular to the streamwise direction (see figure \ref{fig:snapshot} for reference), while the laser sheet is parallel to the $yz$-plane. We note that due to the small stereoscopic angle between the 2 cameras of our PTV system, measurements in the $y$ (streamwise) and $x$ (transverse) directions have a greater spatial redundancy than the $z$ (depth) component. As a consequence, positions recorded in the $z$ component tend to be noisier, leading to slightly less accurate estimates of Lagrangian velocity and acceleration along this component. Therefore, given the global symmetry of our setup, all statistical quantities for the $z$ component will be considered to be identical to those for the $x$ component.

In order to explore the statistical multiscale spatial properties in the present porous medium flow and possible turbulent-like signatures, we analyse the tracers dynamics at the light of the tools typically used in turbulence. In particular, we investigate the two-point statistical properties of velocity fluctuations, defined as $u_i^\prime = u_i - \langle u_i\rangle$, with the average done over all the trajectories obtained with the PTV (one-point statistics can be found in the supplemental material \citep{suppMat}).

\begin{figure}
  \centerline{
    \includegraphics[width=8.8cm,trim={0 0.75cm 0 1.25cm},clip]{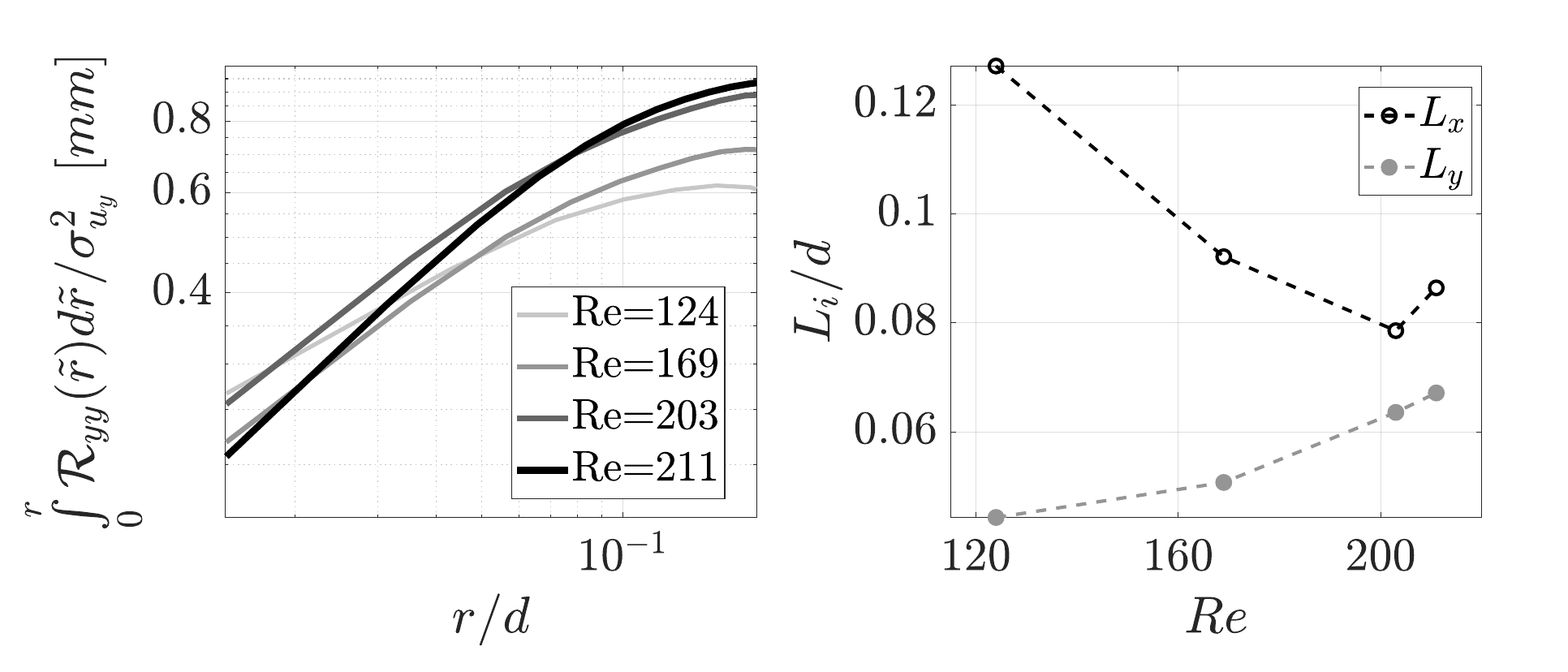}}
  \caption{Left: Cumulative integral of the correlation function $\mathcal{R}$. The plateau allows us to compute $L_y$. Right: Integral correlation lengths calculated for the streamwise ($y$) and transversal ($x$) directions. }
 \label{fig:corrL}
\end{figure}

We first address the large scale properties of the flow and calculate the integral correlation length $L$. To this end, we make use of the Eulerian auto-correlation tensor, $\mathcal{R}_{ij}(r) = \langle u^\prime_i(\mathbf{x}+\mathbf{r})u^\prime_j(\mathbf{x})\rangle$ (details on the streamwise, transversal and crossed correlation functions are shown in~\cite{suppMat}). The correlation lengths in the transversal ($x$) and streamwise ($y$) directions are given by $L_i = \mathrm{lim}_{r\rightarrow \infty}\int_{0}^{r}\mathcal{R}_{ii}(\tilde{r})d\tilde{r}/\sigma _{u_{i}}^2$, with $i=x$ and $i=y$ respectively. Figure~\ref{fig:corrL}(Left) shows -for the streamwise component, but the conclusion holds also for the transversal one- that an asymptotic limit of the cumulative integral in this definition is well converged for all experiments at all the Reynolds numbers we investigated. 

It is worth noting that $L_x$ tends to decrease with $Re$ whereas the opposite trend is observed for $L_y$, as shown in figure~\ref{fig:corrL}(Right). Overall, the large scale correlations of velocity fluctuations are characterized by transverse and streamwise integral scales which are (i) of the order of one tenth of the particle diameter (which is commensurate to the typical pore-scale~\cite{suppMat}), (ii) $Re$-dependent and (iii) with a trend to become isotropic as $Re$ increases (with $L_x/L_y\rightarrow 1$).

In order to zoom-in into the multiscale properties of the fluctuating velocity field, we calculate the Eulerian second-order structure function $S_2(r)$ of streamwise and transversal components. This quantity is of particular interest in turbulence because it carries one of the most celebrated signatures of turbulence at inertial scales, with a characteristic Kolmogorovian $r^{2/3}$ scaling. 

\begin{figure}
  \centerline{
    \includegraphics[width=8.8cm,trim={0.61cm 0.5cm 0.77cm 0.5cm},clip]{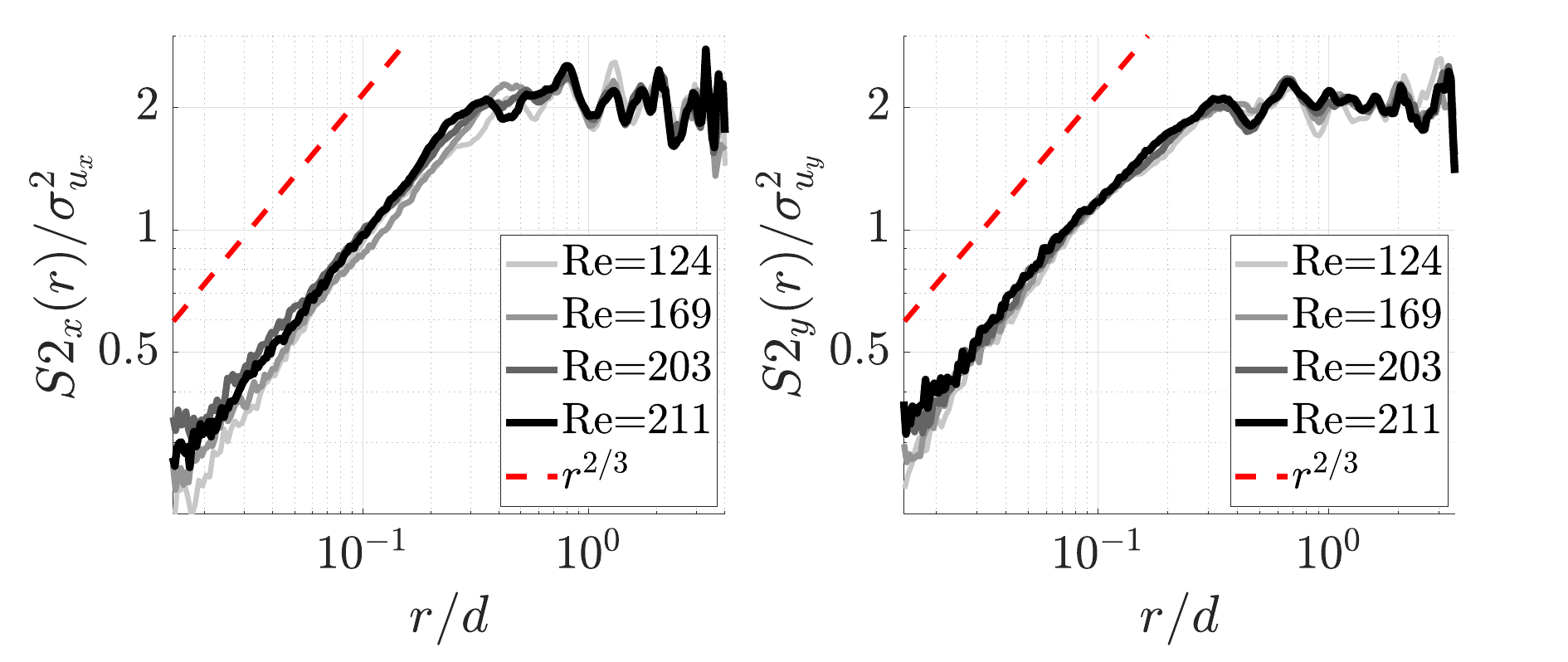}}
  \caption{ Second-order structure functions for each velocity component. A two-thirds scaling (compared in dashed lines) is evident at the small scales $dr<0.2d$.}
 \label{fig:s2}
\end{figure}

Figure \ref{fig:s2} shows $S_2(r)$ calculated for the streamwise component of the velocity $u^\prime_y$ and for the transversal component $u^\prime_x$, $S2_y$ and $S2_x$ respectively. Remarkably, for all the investigated $Re$, they both show a clear $r^{2/3}$ power law over almost one decade of scales, from the smallest resolved scale (of the order of $0.02d$) and up to $r\approx 0.2d$, which is of the order of magnitude of the calculated integral length scale. This reveals a local spatial flow dynamics were energy is distributed across scales in a strikingly similar way as it would in a turbulent flow. At larger scales $S2_{x,y}$ reach a plateau at the asymptotic value of $2\sigma_{u_{x,y}}^2$ as expected for uncorrelated large scale dynamics.

Although highly appealing, the existence of a turbulent-like $r^{2/3}$ power law for $S_2$ is not sufficient to claim the existence of an inertial cascade (where energy is not simply distributed across scales, but actually flows across scales). The existence of an energy cascade is indeed typically related to the Eulerian third-order structure function and the celebrated Kolmogorov's "4/5th law", $S_3^\parallel(r) = \frac{4}{5}\epsilon r$~\cite{frisch1995} (with $S_3^\parallel$ the longitudinal third order structure function), characteristic of a direct cascade of energy (where energy flows from large to small scales at the rate of $\epsilon$ per unit mass and unit time) as observed in 3D turbulent flows. Given the Lagrangian nature of the original dataset recorded in our setup, we use here a mathematically equivalent relation, based on the crossed velocity - acceleration structure function $S_{au}(r)=\langle \delta \mathbf{a}\cdot \delta \mathbf{u} \rangle$. Indeed, it can be shown that for a locally homogeneous and isotropic turbulent flow~\citep{Mann1999,Ott2000,HillRJ2006}
\begin{equation}
     S_{au}(r)=\langle \delta \mathbf{a}\cdot \delta \mathbf{u} \rangle = - 2\epsilon.
     \label{eq:dadu}
\end{equation}
\noindent In this relation the \textit{minus} sign is characteristic of a direct energy cascade. 
$S_{au}(r)$ therefore yields information about the energy cascade by (i) its sign and (ii) its absolute value $2\epsilon$, which is expected to be constant across inertial scales and to give a direct estimate of the energy transfer rate $\epsilon$ (which in stationary conditions equals the energy injection and the energy dissipation rate). 

We calculated $S_{au}$ from the $y$ and $x$ components only, assuming perfect isotropy between the transversal coordinates $x$ and $z$, which is a reasonable assumption given the cylindrical symmetry of our experiment. We therefore estimate $\langle \delta \mathbf{a} \cdot \delta \mathbf{u}\rangle=2\langle \delta a_x \delta u_x\rangle +\langle \delta a_y\delta u_y\rangle$. 

\begin{figure}
    \includegraphics[width=\linewidth,trim={0.75cm 0.35cm 0.55cm 1.2cm},clip]{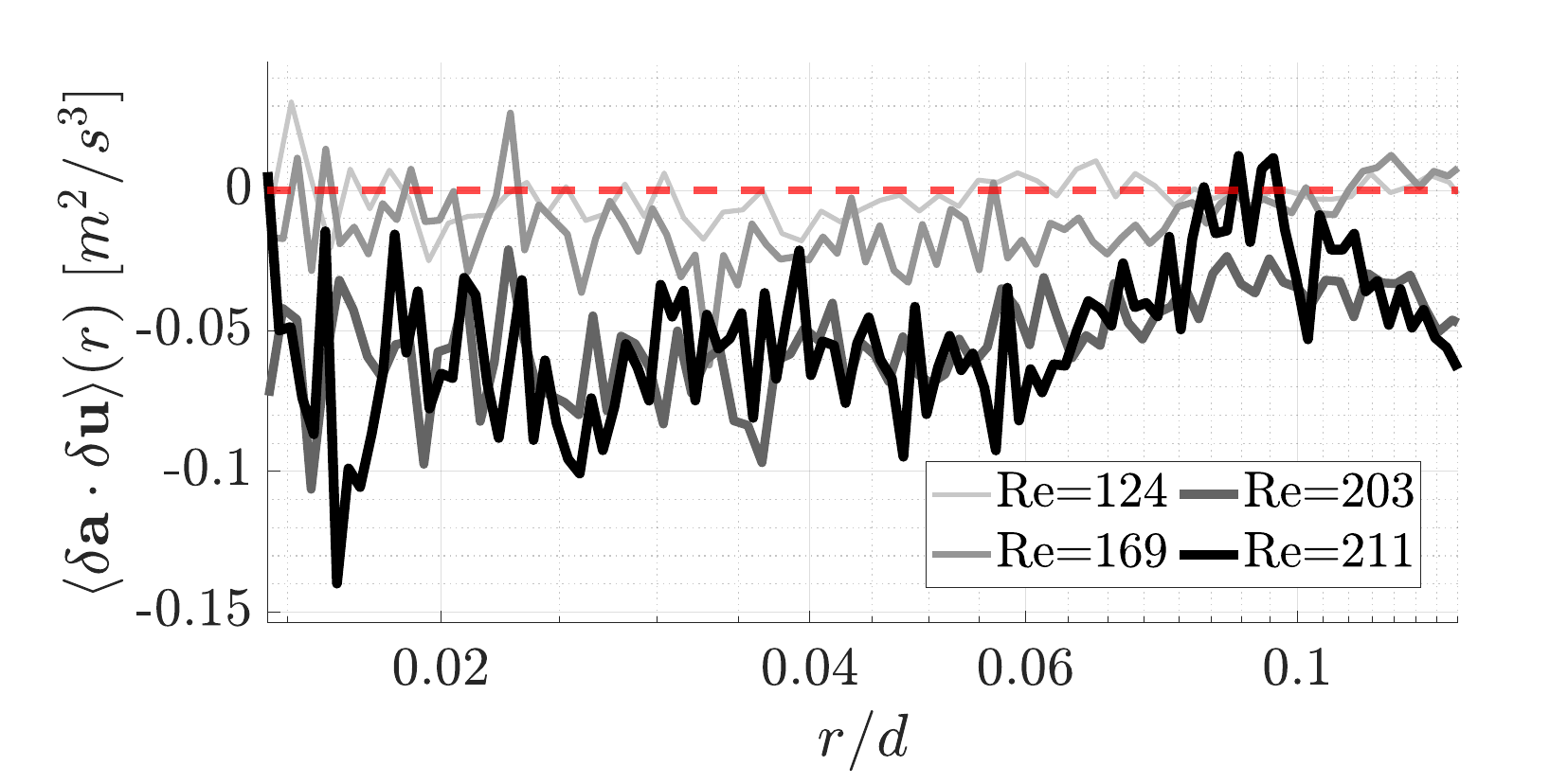}
    \includegraphics[width=0.75\linewidth,trim={0.15cm 0.15cm 0.15cm 0.85cm},clip]{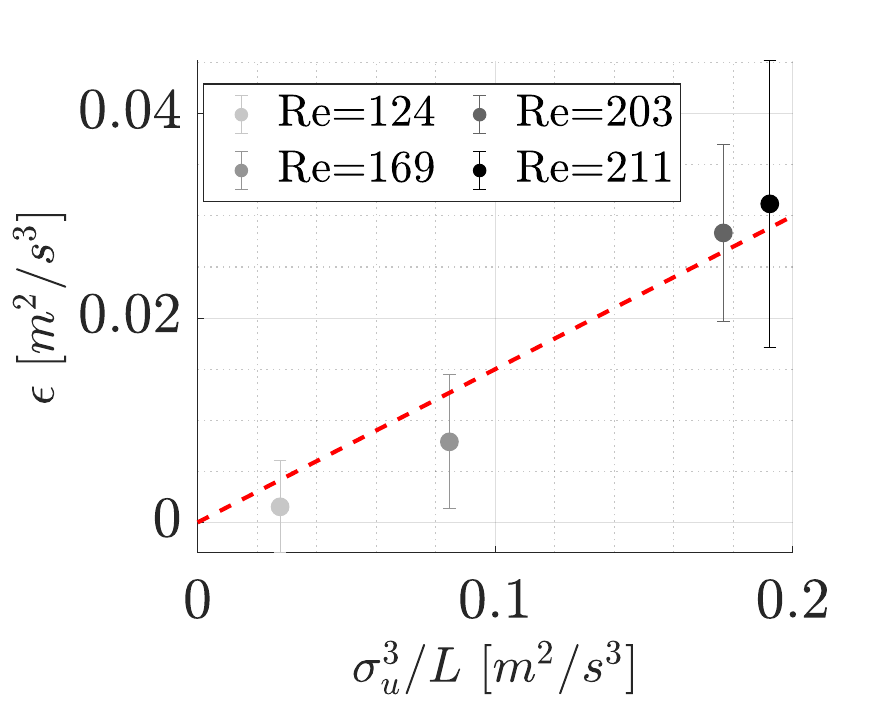}
\caption{Top: The acceleration-velocity structure function, which is supposed to remain constant in a homogeneous and isotropic turbulent flow, and its negative sign indicates the direction of the energy cascade. Bottom: Averaged energy dissipation rate as a function of $\sigma_u^3$}

 \label{fig:dadu}
\end{figure}

Figure \ref{fig:dadu}(Top) shows the results for the experiments at different $Re$. It is first observed that in spite of some scatter $S_{au}$ remains relatively constant over the range of scales $r$ where the $r^{2/3}$ scaling was observed for $S_2$. Besides, it keeps a persistent negative sign in that range, especially in the higher $Re$ flows. In the spirit of classical turbulence, this would be associated to the presence of a direct energy cascade at inertial scales.

Using relation~(\ref{eq:dadu}) as a formal analogy to the turbulent energy cascade, we can calculate the value of the equivalent rate of energy transfer across scales $\epsilon$ from the mean value of the plateau of $S_{au}$, so that $\epsilon = |\overline{\langle S_{au}\rangle}|/2$ (in this case, the second average $\overline{ \ \cdot \ }$ is done over $r$). From this estimate, we can compute several important parameters classically used to characterize of turbulence by

\begin{itemize}
 \item[(i)] \noindent exploring at large scales the classical turbulent relation between fluctuating velocity $\sigma_u$, energy injection rate $\epsilon$ and integral scale $L$:

\begin{equation}
    \epsilon = \frac{C_\epsilon \sigma_u^3}{L},
    \label{eq:u3_vs_eps}
\end{equation}
\noindent where $C_\epsilon$ is a non-universal constant typically in the range $C_\epsilon\in[0.5;1]$ in turbulence~\citep{vassilicos2015}. Subsequently, the value of $C_\epsilon$ then allows to estimate the ``Taylor based Reynolds number'', $R_\lambda$, of the flow, which is commonly used to characterize the intensity of turbulence.
 \item[(ii)] \noindent Determining the equivalent Kolmogorov -or dissipation- scale of the flow $\eta=(\nu^3/\epsilon)^{1/4}$.
 \item[(iii)] \noindent Determining the equivalent of the Kolmogorov constant $C_2$ for the second order structure function $S_2(r) = C_2 (\epsilon r)^{2/3}$. 
\end{itemize}

Regarding the first point, Figure~\ref{fig:dadu}(Bottom) shows the energy transfer rate $\epsilon$ estimated from the energy cascade relation~\ref{eq:dadu} at inertial scales as a function of $\sigma_u^3/L$ (here we defined $\sigma_u=\sqrt{\frac{2\sigma_{u_x}^2+\sigma_{u_y}^2}{3}}$ and $L=1/3(2L_x+L_y)$). Within errorbars, the trend is found to follow the same relation~\eqref{eq:u3_vs_eps} as in classical turbulence, with $C_\epsilon = 0.15\pm 0.05$.

We can then estimate $R_\lambda = \sqrt{\sigma_u L /\nu C_\epsilon}$~\cite{Pope2000}. The corresponding values are shown in table \ref{tab:table1}. As it could be expected, $R_\lambda$ increases with $Re$, although $R_\lambda$ increases only 15\% when $Re$ increases more than 50\%. In classical turbulence, such values ($R_\lambda\approx 80$) would be qualified of moderately turbulent, and are for instance produced in wind tunnel experiments.  The small variation observed for $R_\lambda$ suggests that the turbulence that develop in the porous medium, weakly depends on the superficial velocity driving the flow.

The energy injection rate also allows us to estimate the expected low bound of the inertial range, namely the Kolmogorov length scale $\eta = (\nu^3/\epsilon)^{1/4}$. The values obtained are shown in table \ref{tab:table1}. As it could be expected, they decrease when $Re$ and $\epsilon$ increase. This indicates that smaller scales emerge as the energy injection and the turbulence increase, following the classical phenomenology of the turbulent energy cascade. 

Finally, by fitting the inertial scaling $S_2(r) = C_2 (\epsilon r)^{2/3}$, the estimate of $\epsilon$ allows to determine the equivalent of the Kolomogorov constant $C_2$ for the present turbulent-like dynamics (see~\cite{suppMat}). Note that we define here $C_2$ based on the total second order structure function. The corresponding values are given in table~\ref{tab:table1}, showing that, within errorbars, a unique value of $C_2=3.1\pm 0.3$ reasonably describes the inertial range energy distribution for all values of the superficial Reynolds number explored. This value is about 2.5 times smaller than the generally accepted value for homogeneous isotropic fluid turbulence (for which $C_2=11C_2^\parallel/3$, with $C_2^\parallel\approx 2.1$ is the Kolmogorov constant for the longitudinal second order structure function).

\begin{table}
\begin{center}
  \begin{tabular}{|c|c|c|c|c|}
    \hline
   $Re$ & $\epsilon$ [m$^2$.s$^{-3}$] & $\eta\ [\mu]m$ & $R_\lambda$ & $C_2$ \\
 \hline
    124 & $(1.50\pm0.45)\text{e-}03$ & $161\pm 12.1$  & 71 & $3.4\pm 0.3$\\
    169 & $(7.90\pm0.66)\text{e-}03$ & $106\pm 22.2$  & 73 & $2.8 \pm 0.3$ \\ 
    203 & $(2.83\pm0.87)\text{e-}02$ & $77.1\pm 5.93$  & 80 & $3.0 \pm 0.3$ \\ 
    211 & $(3.12\pm1.40)\text{e-}02$ & $75.2\pm 8.44$  & 84 & $3.2 \pm 0.3$ \\
    \hline
  \end{tabular}
  \caption{Different turbulent parameters calculated in terms of the energy injection rate $\epsilon$.}  

  \label{tab:table1}
\end{center}
\end{table}

To summarize, We have investigated the flow developing in a transitional fixed bed of hydrogel beads and found that, even though the flow is globally stationary, it develops a spatial multiscale dynamics which shares striking analogies with classical fluid turbulence. 

In particular, we have shown the existence of a direct energy cascade, with a typical energy injection scale $L$ commensurate with the pore scale and a characteristic energy transfer rate across scales $\epsilon$. At the large scales the classical turbulent relation $\epsilon = C_\epsilon \sigma_u^3/L$ is verified. At inertial scales, the energy distribution follows the classical Kolmogorovian scaling $S_2 = C_2 (\epsilon r)^{2/3}$. Although the claim of their universality will require further experiments, both parameters $C_\epsilon$ and $C_2$ where found to be reasonably constant, hence giving a consistent parametrization of fluctuations at large and inertial scales over the range of Reynolds numbers explored here.

These results open several interesting perspectives, for porous media physics and beyond. First, regarding porous media, the analogy with turbulence may help building new approaches for their mixing and global transport properties. Indeed, the capacity of turbulent flows to disperse substances and fields is intimately related to the inertial multiscale dynamics~\cite{bib:bourgoin2015_JFM}.

Secondly, our results show that, to some extent, such transitional porous media flows can be considered as a true experimental model of frozen turbulence~\cite{Schlipf2011} where the underlying velocity field does not vary in time but has a rich spatial multiscale dynamics. As such, this system may help disentangling the intricate role of spatial and temporal fluctuations occurring in real turbulence, and its impact on the energy cascade and subtle phenomena such as pair dispersion and small-scale intermittency.

Finally, our findings enrich the class of out-of-equilibrium systems exhibiting a Kolmogorovian-like energy cascade, in a comparable way to what was recently reported for active matter~\cite{Bourgoin2020}. Building such bridges between apparently disconnected phenomena for which no generic theoretical framework has yet emerged may be crucial to identify the key common ingredients for their understanding.

\bibliography{bibliography.bib}

\clearpage
\onecolumngrid
\setcounter{figure}{0}    

\section*{Supplemental Material} 

\section{Stationarity}

\begin{figure}[h]
	\centerline{
  	\includegraphics[width=.9\textwidth]{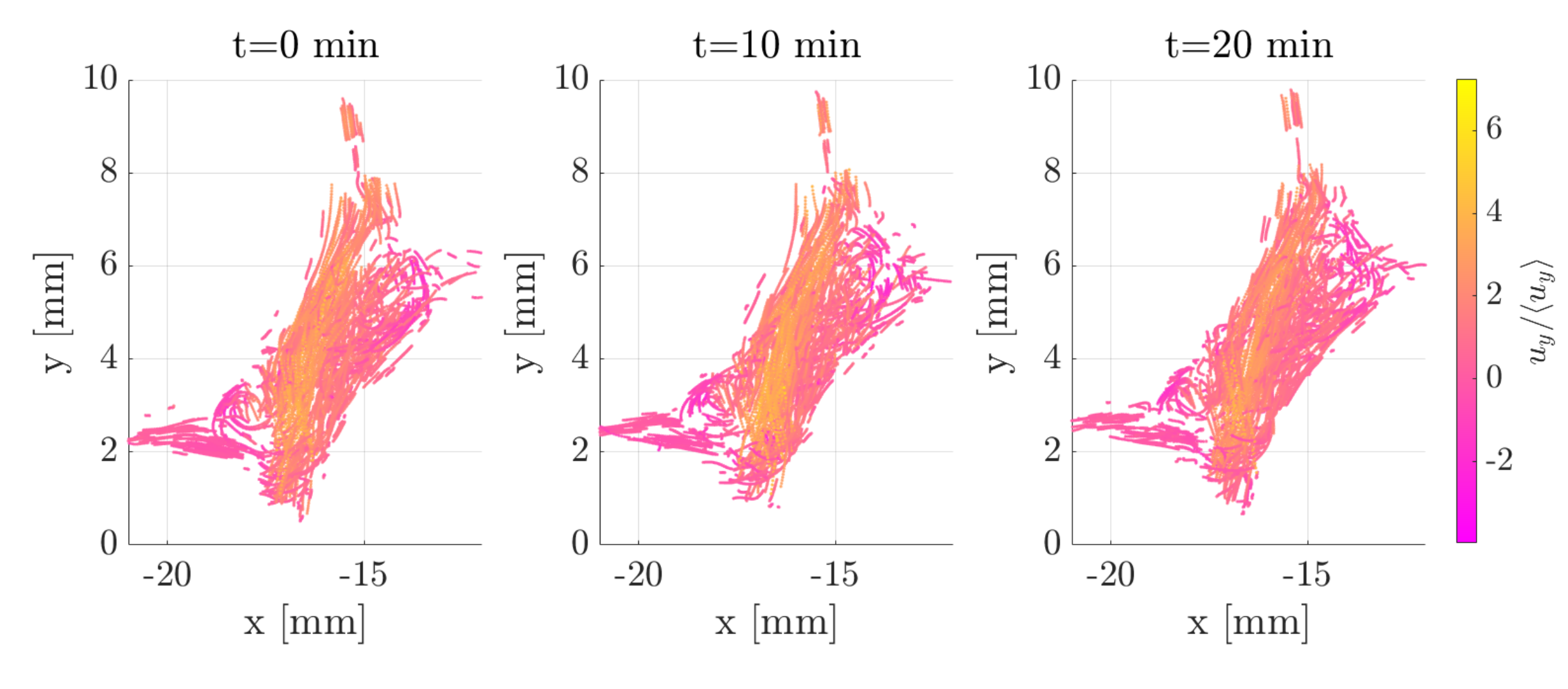}}
      \caption{\small 2-second timelapse of the Lagrangian velocity field inside a generic pore. The measurements were started 10 minutes apart, and the velocity field does not significantly change in time.}      
	\label{fig:stationarity}
\end{figure}

\noindent In order to study stationarity, we computed the Lagrangian velocity field and visualized how the saturating flow evolves in time. Figure \ref{fig:stationarity} shows three 2 minute time-lapsed individual measurements of the flow passing through the same bed. The measurements were started ten minutes apart from each other ($t=[0,10,20]min$), and the figure shows a particular pore. The limited white sections correspond to four different surrounding beads. It can be seen that the flow indeed does not change in time. For comparison, the characteristic time of the integral length $T_L$ defined in terms of $\sigma_u$ and $L$, is $L/\sigma_u = [0.4,0.2,0.2,0.2]$ for the Reynolds number based on the superficial velocity $Re=[124,169,203,2011]$ respectively. 

\section{One-point statistics}

\begin{figure}
  \centerline{
    \includegraphics[width=0.8\textwidth]{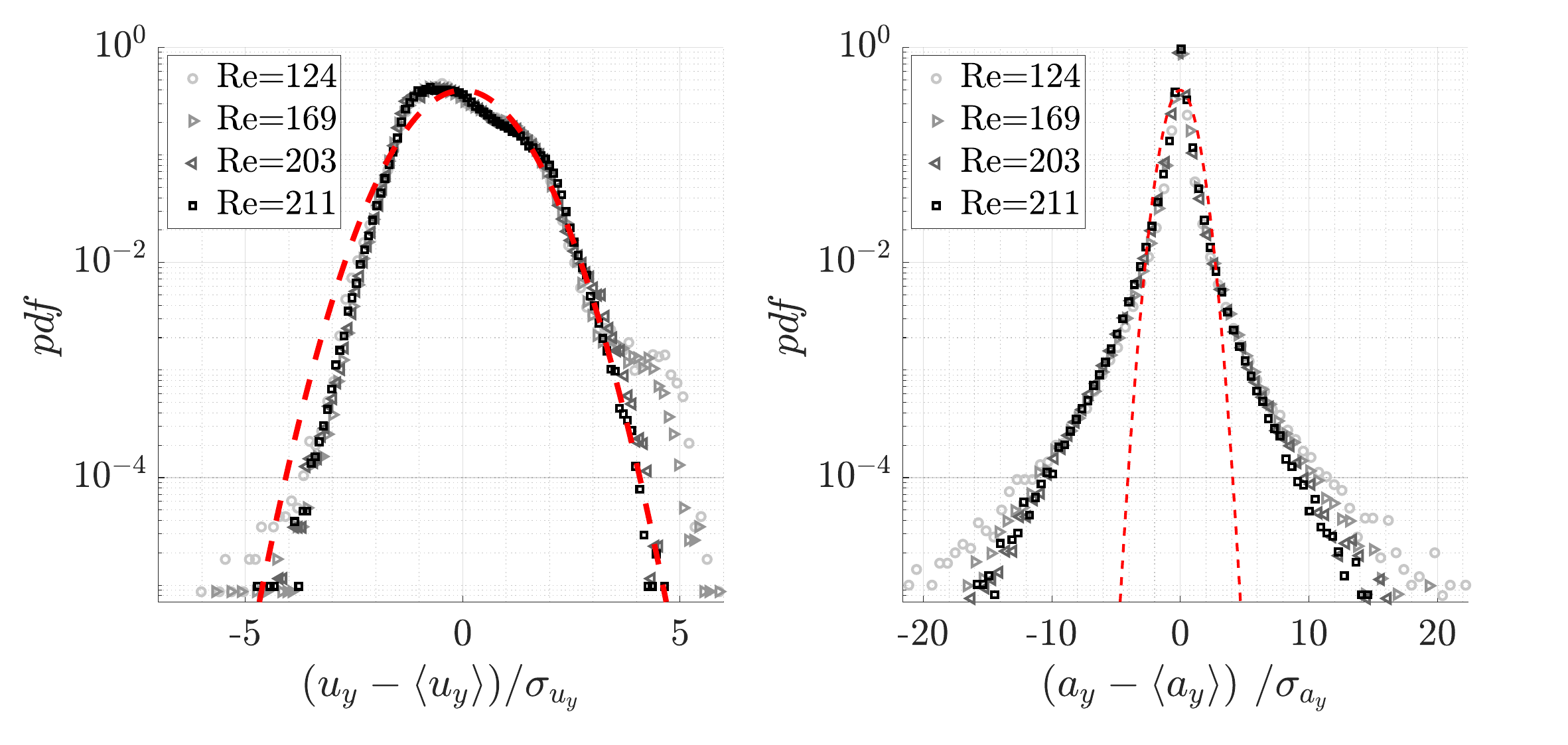}}
  \caption{Probability density function (pdf) the streamwise components of the velocity and acceleration (Left and right respectively). They are expressed in such a way so that they can be compared to a Gaussian distribution of mean zero and standard deviation equal to one. Both distributions present exponential tails, which are observed in turbulent flows where extreme values of acceleration are probable.}
 \label{fig:accPDF}
\end{figure}

\begin{figure}[h]
	\centerline{
  	\includegraphics[width=0.8\textwidth]{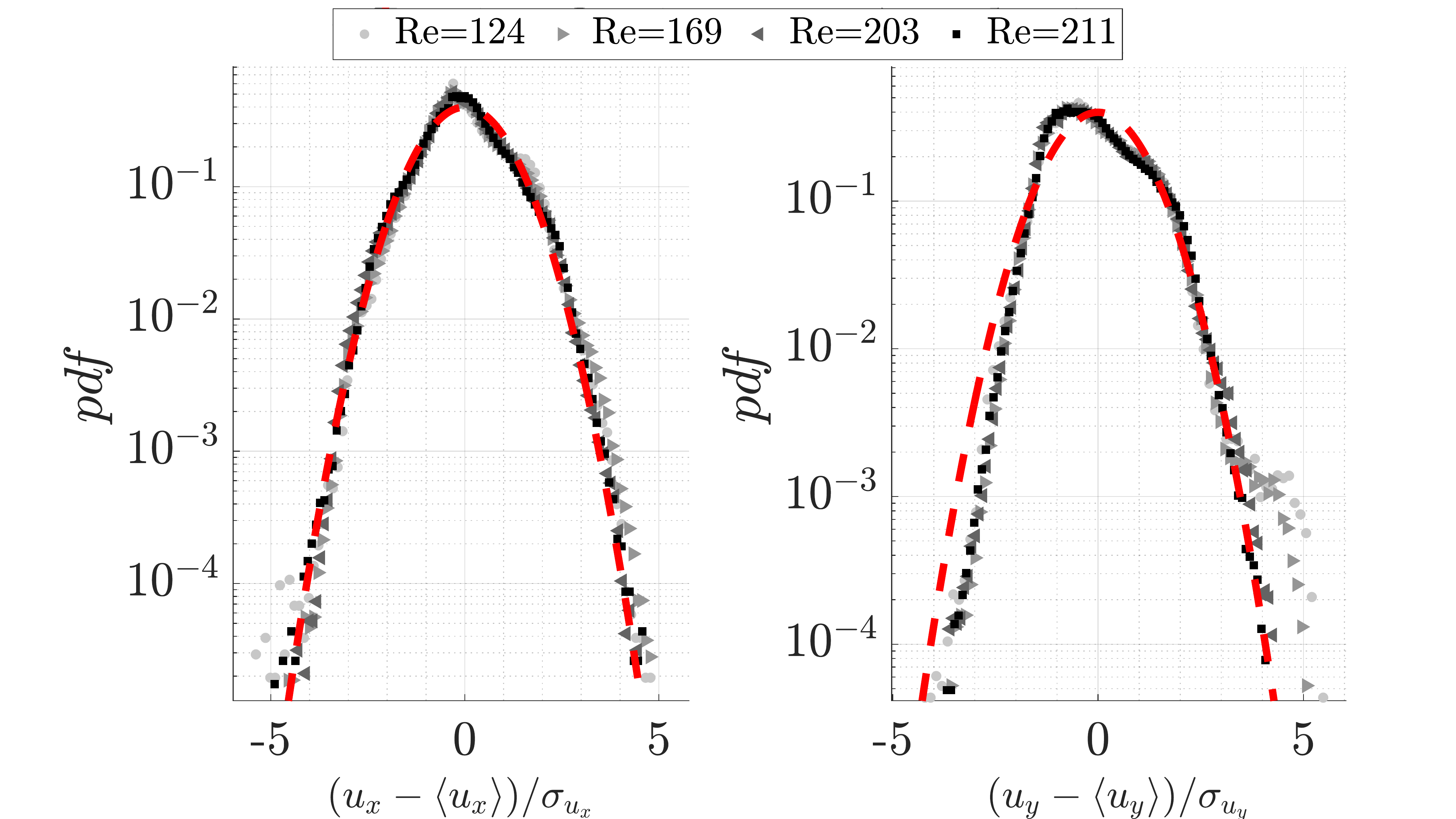}}
      \caption{\small Transversal and streamwise velocity probability density function for the different flows studied (Left and right, respectively). The pdf's are normalized and compared to a Gaussian distribution with zero mean and a standard deviation equal to one (shown in red dashed lines), so as to visualize their symmetry around zero.}      
	\label{fig:pdf_components}
\end{figure}

Figure \ref{fig:accPDF} shows the centered (by the mean) and reduced (by the standard deviation) probability density functions (pdf) for the streamwise component of the velocity and acceleration, $u_y$ and $a_y$ respectively, for the different Reynolds numbers considered. Table \ref{tab:table0} shows the values of the first two statistical moments of the streamwise components of velocity and acceleration. The velocity pdf is close to Gaussian although it is slightly skewed towards positive values. In contrast, the pdfs of the transversal components (figure \ref{fig:pdf_components}) are symmetric around their mean value.

Such behavior, with skewed streamwise fluctuations of velocity and symmetric transverse fluctuations, have also been observed in \cite{Souzy2020, Patil2013} for experiments at smaller Reynolds numbers, although the skewness reported in these studies is significantly more pronounced than what we observe in this Letter.
The origin of the streamwise skewness might be explained by the vertical pressure difference driving the upward flow in the $y$ direction, susceptible to promote positive velocity events, while the reduced skewness in the present studies very likely reveals the richer taxonomy of spatial structures (such as recirculations) in the higher Reynolds number regime explored here.  

On the other hand, the acceleration pdfs are remarkably similar to those observed in fully turbulent flows \citep{LaPorta2001,Voth2001}, where extreme values of acceleration can be reached, as it is evidenced by the highly non-Gaussian stretched tails. Pdfs with similar turbulent-like characteristics have also been reported in previous porous medium flows \cite{Holzner2015} at lower Reynolds numbers ($Re\sim 0.4$). These observations are striking as the porous medium flow fields considered here and in~\cite{Holzner2015} are stationary in time, contrary to fluid turbulence. However, they share the existence of tortuous structures within the flow, with velocities along Lagrangian paths that vary significantly as fluid particles move from regions of high and low porosity. As a result, the fluctuations of Lagrangian acceleration in the present steady (though spatially complex) flow field appear to develop appealing similarities to what is observed in actual turbulence.

\begin{table}
\begin{center}
 \begin{tabular}{|c|c|c|c|c|c|}
    \hline
   $Re$ & $U$ & $\langle u_y \rangle $ &  $\sigma _{u_y}$  & $\langle a_y \rangle $  & $\sigma _{a_y}$ \\
 \hline
    124 & $8.92$ & $38.77$ &   34.80 &  $-87.23$ & 5.19e3  \\
    169 & $12.13$ & $51.34$ & $47.96$ & $-134.68$ &  $6.83$e3 \\ 
    203 & $14.57$ & $60.06$ & 61.70 & $-260.68$ & 8.85e3 \\ 
    211 & $15.17$ & $61.81$ & 64.73 & $-276.88$ & 1.01e4 \\
    \hline
  \end{tabular}
  \caption{Mean and standard deviation values of the streamwise components of the velocity and acceleration. All quantities are shown in millimeters.}  
\label{tab:table0}
\end{center}
\end{table}

\clearpage
\newpage

\section{Correlation functions}
\begin{figure}[h]
	\centerline{
  	\includegraphics[width=0.8\textwidth]{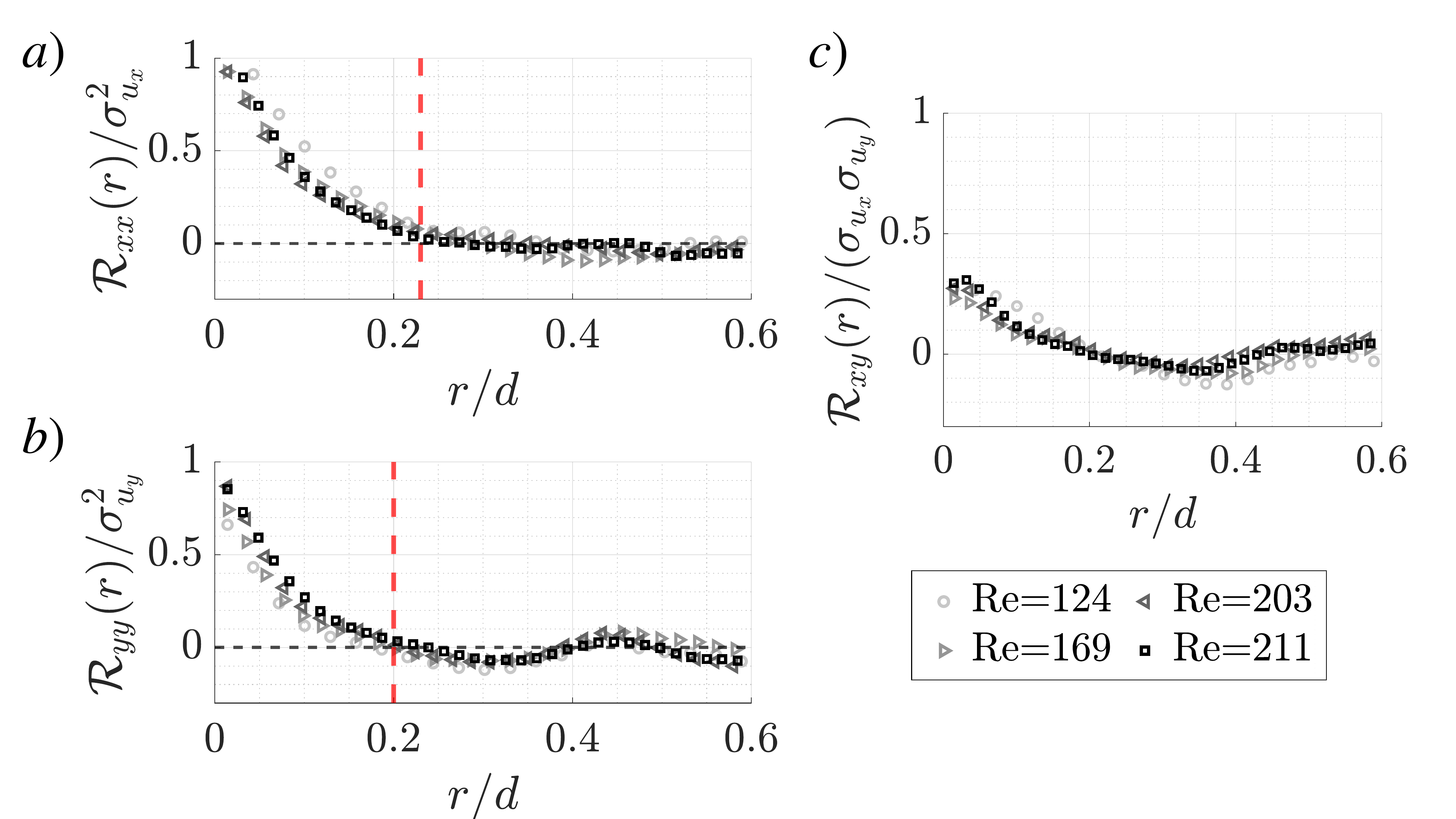}}
      \caption{\small Transversal and longitudinal structure functions (a) and b) respectively), normalized by their respective standard deviations. For reference, a horizontal line at zero is marked with a black dashed line and the vertical red dashed line shows the approximate point where $\mathcal{R}$ crosses zero. Subfigure c) shows the crossed correlation function.}      
	\label{fig:correlation}
      \end{figure}

\noindent In order to compute the correlation lengths, the correlation functions $\mathcal{R}$ are needed. They are defined as
\begin{equation*}
    \mathcal{R}_{ij} = \langle  u^\prime_i (\mathbf{x}+\mathbf{r})  u^\prime_j (\mathbf{x}) \rangle.
\end{equation*}
\noindent They tend to one when $r=0$ and two elements of fluid are no longer correlated at distances $r$ when $\mathcal{R}_{ij}=0$, which is the correlation length $L$. The computed correlation functions are shown in figure \ref{fig:correlation}, and a dashed black line is shown at $\mathcal{R}_{ij}=0$ for visualization purposed. Both $\mathcal{R}_{xx}$ and $\mathcal{R}_{yy}$ tend to one at $r=0$ and become zero at different correlation lengths that depend on the Reynolds number and the component of the velocity that is being taken into account (see figure 4 in the Letter). Once they cross zero a slight oscillating pattern is observed, which is due to the presence of the beads. This shows that there is still a slight correlation present that oscillates spatially.

The crossed correlation function (fig. \ref{fig:correlation}\textit{c)}) shows that $u_x$ and $u_y$ are slightly correlated at $r=0$ ($\mathcal{R}_{xy}(r=0)\approx 0.25$) and the correlation remains low for all separation distances.

\section{Typical pore length}

\begin{figure}[h]
	\centerline{
  	\includegraphics[width=0.33\textwidth]{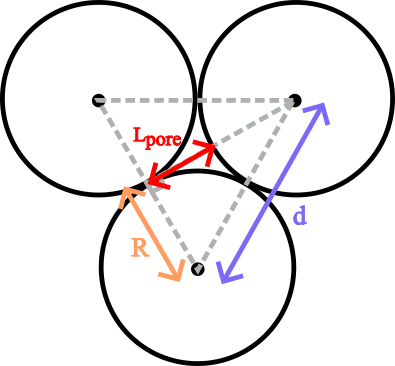}}
      \caption{\small Geometrical estimate of the typical pore length. $d$ is the sphere diameter, $R$ its radius and $L_{pore}$ the typical pore scale. $L_{pore}$ can be calculated using Pythagoras's theorem.}      
	\label{fig:pore_estimate}
\end{figure}

Here we provide a rough estimate of the typical pore length, by considering simple geometrical arguments. Let us have three spheres of radius $d/2$ closely packed together, as shown in figure \ref{fig:pore_estimate}. The three centers are joined by a triangle, and the typical pore length is shown in the plot as $\ell_{pore}$. The other relevant lengths are also shown as $d$ (the sphere diameter) and $R=d/2$ the sphere radius. By making use of Pythagoras's theorem, we have that $h = d/\sqrt{2}$. Considering that $L_p = d/\sqrt{2}-d/2$, we have that a typical pore is $L_{pore}\approx 0.2d$, which is of the order of magnitude of the computed correlation lengths. It is also consistent with figures \ref{fig:correlation}\textit{a)} and \textit{b)}, at the point where the correlation becomes zero.

\newpage
\section{Kolmogorov's constant}

\begin{figure}[h]
	\centerline{
  	\includegraphics[width=.33\textwidth]{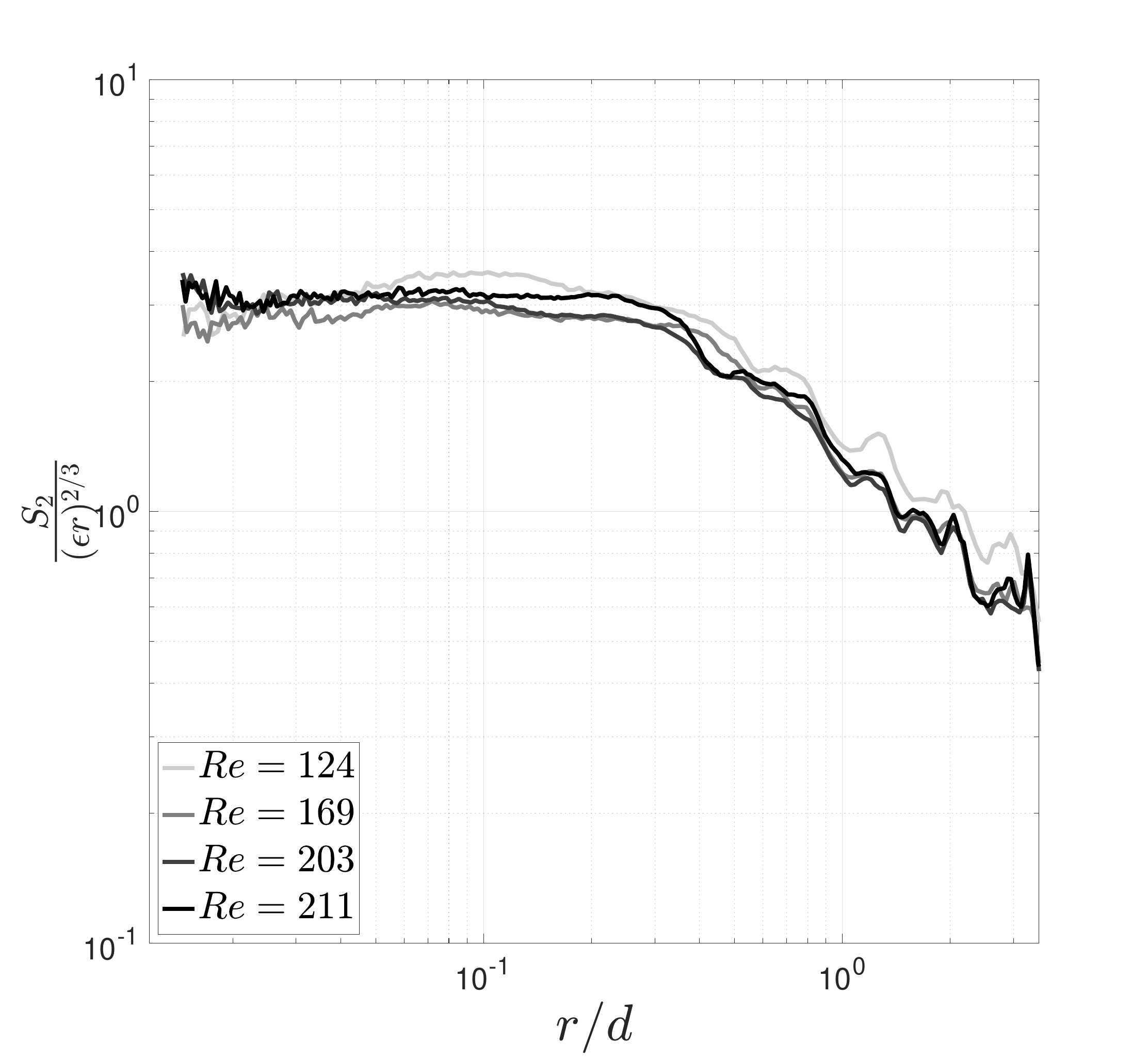}
\hspace{.12\textwidth}

    \includegraphics[width=.33\textwidth]{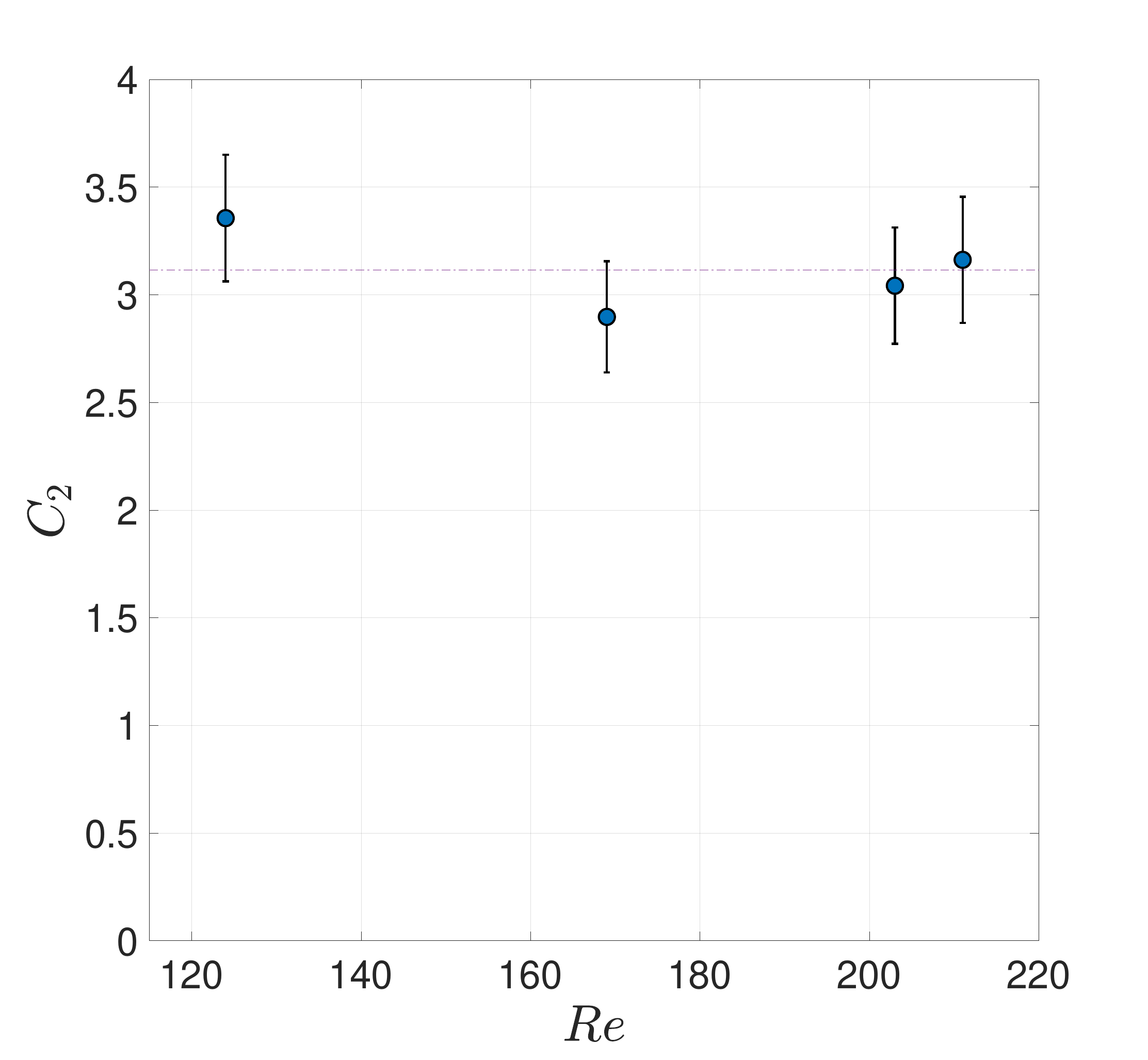}}
   
      \caption{\small (Left) Compensated total second structure function $S_2(r)/(\epsilon r)^{2/3}$. (Right) Estimated value of the Kolmogorov constant $C_2$ (the dashed represents the mean of all four values of $C_2$).}      
	\label{fig:S2Comp}
      \end{figure}

The Kolmogorov constant $C_2$ can be estimated by fitting the power law scaling for $S_2$ at inertial scales. To do so, we follow here the classical method used in turbulence, which consists in plotting the compensated structure function $S_2(r)/(\epsilon r)^{2/3}$. This requires to know \textit{a priori} the value of the energy dissipation rate, which has been obtained here from the inertial range value of the crossed velocity-acceleration structure function $S_{au}$. Figure~\ref{fig:S2Comp}(Left) shows the compensated structure function for the different experiments (at different superficial velocities) explored here. The inertial range scaling appears as a plateau giving the value of $C_2$. Note that we consider here the total structure function $S_2=(2S_{2x} + S_{2y})/3$. Figure~\ref{fig:S2Comp}(Right) shows the value of $C_2$, estimated as the average of the plateau over the range of scales $r/d < 0.1$ ; the errorbars correspond to the standard deviation of the plateau over the same range.

\end{document}